\documentclass[a4paper,fleqn
]{cas-dc}

\usepackage[numbers]{natbib}
\usepackage{enumitem}
\usepackage[utf8]{inputenc}
\usepackage{algorithm}
\usepackage{float}
\usepackage{algpseudocode}
\def\tsc#1{\csdef{#1}{\textsc{\lowercase{#1}}\xspace}}
\tsc{WGM}
\tsc{QE}
\tsc{EP}
\tsc{PMS}
\tsc{BEC}
\tsc{DE}

\begin{document}
\let\WriteBookmarks\relax
\def\floatpagepagefraction{1}
\def\textpagefraction{.001}

\shorttitle{OCU-Net Model for Accurate Oral Cancer Detection}

\shortauthors{Albishri et al.}

\title [mode = title]{OCU-Net: A Novel U-Net Architecture for Enhanced Oral Cancer Segmentation}





%
\author[1,2]{Ahmed Albishri}[
                        orcid=0000-0001-6002-5734]



\ead{aa8w2@umsystem.edu}



\affiliation[1]{organization={Computer Science, School of Science and Engineering,  University of Missouri-Kansas City},
    city={Kansas City},
    country={USA}}
\affiliation[2]{organization={College of Computing and Informatics, Saudi Electronic University},
    city={Riyadh},
    country={Saudi Arabia}}
    
\affiliation[3]{organization={Department of Oral and Craniofacial Sciences, School of Dentistry, University of Missouri - Kansas City},
    city={Kansas City},
    country={USA}}

\author[1]{Syed Jawad Shah}


\author[1]{Yugyung Lee}[%
   ]



\author[3]{Rong Wang}




\credit{Conceptualization and design of the study, Data curation, Writing - Original draft preparation, Review and editing, Funding acquisition}







\begin{abstract}
Accurate detection of oral cancer is crucial for improving patient outcomes. However, the field faces two key challenges: the scarcity of deep learning-based image segmentation research specifically targeting oral cancer and the lack of annotated data. 
Our study proposes OCU-Net, a pioneering U-Net image segmentation architecture exclusively designed to detect oral cancer in hematoxylin and eosin (H\&E) stained image datasets. OCU-Net incorporates advanced deep learning modules, such as the Channel and Spatial Attention Fusion (CSAF) module, a novel and innovative feature that emphasizes important channel and spatial areas in H\&E images while exploring contextual information. In addition, OCU-Net integrates other innovative components such as Squeeze-and-Excite (SE) attention module, Atrous Spatial Pyramid Pooling (ASPP) module, residual blocks, and multi-scale fusion. The incorporation of these modules showed superior performance for oral cancer segmentation for two datasets used in this research. Furthermore, we effectively utilized the efficient ImageNet pre-trained MobileNet-V2 model as a backbone of our OCU-Net to create OCU-Net$^m$, an enhanced version achieving state-of-the-art results.  
Comprehensive evaluation demonstrates that OCU-Net and OCU-Net$^m$ outperformed existing segmentation methods, highlighting their precision in identifying cancer cells in H\&E images from OCDC and ORCA datasets. 
\end{abstract}




\maketitle

\section{Introduction}
Oral cancer is a significant global health burden, with more than 377,000 new patients and over 177,000 deaths annually worldwide \citep{sung2021global}. Oral squamous cell carcinoma (OSCC) originates from the mucosal lining of the oral cavity and accounts for more than 90\% of oral cancer cases  \citep{warnakulasuriya2009global}. The 5-year survival rate of oral cancer varies from 20\% to 80\% depending on the stage of diagnosis \citep{aghiorghiesei2022world}. Accurate diagnosis is crucial for treatment planning and management. Currently, biopsy followed by histological assessment of hematoxylin and eosin (H\&E) stained tissue specimens is the gold standard for oral cancer diagnosis \citep{Speight2018pathology}. However, this process is time-consuming, qualitative, and subjective, with considerable variabilities \citep{wang2021fourier}. Digital pathology-based methods have emerged for quantitative pathological evaluations \citep{baxi2022digital}, with digital whole-slide images (WSIs) used for computer-aided diagnosis (CAD). Developing a reliable and robust automatic cancer tissue segmentation algorithm can significantly contribute to CAD processes.

Artificial intelligence and machine learning have proven useful for making clinical workflows more efficient. Deep learning 
can extract features and has been effective in various image analysis tasks \citep{bengio2017deep, shen2017deep, schwendicke2019convolutional}. Recent studies have shown the potential of deep learning models in enhancing precision medicine for oral cancer management \citep{alabi2022deep}. For instance, deep learning methods have been applied to positron emission tomography (PET) and computerized tomography (CT) images for predicting patient survival and disease-free survival with high accuracy \citep{ fujima2020deep, ariji2020ct}. However, the use of deep learning in oral cancer histopathological image analysis has been limited due to various challenges. Deep learning-based image segmentation methods for oral cancer WSIs are notably more limited than classification, as pixel-level detection is more challenging \citep{manhas2021review}. Also, factors such as data annotation, external validation, comparison with human experts, and increasing multidimensional data availability must be considered to implement deep learning models effectively.

In this study, we introduce the Oral Cancer U-Net Segmentation Models, OCU-Net and OCU-Net$^m$, which are advanced U-Net-based models designed for precise oral cancer segmentation. These models address challenges such as limited research in deep learning image segmentation and the scarcity of annotated data. They achieve state-of-the-art performance in oral cancer detection through image segmentation on H\&E stained image datasets.

The notable innovations of the OCU-Net are (1) the specific design of a novel attention model called Channel and Spatial Attention Fusion (CSAF) and its unique incorporation alongside other deep learning building blocks, including Squeeze-and-Excitation (SE) attention blocks \citep{hu2018squeeze}, the ASPP module with parallel atrous convolutions \citep{chen2017deeplab}, residual blocks \citep{he2015deep}, and multi-scale fusion to enhance oral cancer identification performance; and (2) OCU-Net$^m$ leveraging a pre-trained ImageNet \cite{deng2009imagenet} model, MobileNet-V2 \citep{singh2019shunt}, as the network encoder and feature extractor, enabling it to capture diverse cancer presentations within oral cancer datasets using a scaled model network, despite the limited availability of annotated data.

Our research aims to address the gap in the literature regarding the optimal U-Net architecture for oral cancer detection using deep image segmentation techniques and the scarcity of H\&E image datasets on oral cancer. Our contributions are summarized as follows:

\begin{itemize}
\item First, we proposed two advanced novel architectures, OCU-Net and OCU-Net$^m$, for accurate oral cancer segmentation.
\item Second, we designed a novel Channel and Spatial Attention Fusion (CSAF) module that highlights significant channel and spatial regions while examining contextual information.
\item Third, OCU-Net$^m$ utilized an efficient pre-trained ImageNet \cite{deng2009imagenet} model, MobileNet-V2 \citep{singh2019shunt}, as a network backbone to identify diverse cancer presentations and further enhance the performance of oral cancer segmentation.
\item Lastly, we conducted extensive evaluations using baseline algorithms and ablation studies to demonstrate our model's state-of-the-art performance for oral cancer segmentation on H\&E images from two datasets, OCDC \cite{dos2021automated} and ORCA \cite{martino2020deep}.
\end{itemize}

Our research positions OCU-Net as a valuable tool in clinical settings, enabling earlier and more accurate oral cancer detection.

\section{Related Work}

\subsection{Deep Learning Segmentation}

Deep learning techniques are powerful tools in medical image analysis, facilitating accurate diagnoses and treatment planning across various modalities. For instance, deep learning methods have been employed in CT scans for lung segmentation and identifying COVID-19 infections \cite{albishri2022tlu}. Similarly, in MRI, U-Net-based methods have been proposed for automatic human brain claustrum segmentation \cite{albishri2022unet}. In dental imaging, deep learning techniques have been applied to panoramic radiograph X-rays for tooth segmentation and identification \cite{chandrashekar2022collaborative}. These advancements are transforming medical image analysis and contributing to more accurate diagnoses and personalized patient treatment plans.

FABnet is a deep learning-based approach for simultaneous segmentation of microvessels and nerves in H\&E-stained histology images \citep{fraz2020fabnet}. Musulin et al. \cite{musulin2021enhanced} proposed a two-stage AI-based system for diagnosing oral OSCC using histopathology images. They achieved satisfactory results for multiclass grading and segmentation. Senousy et al. \cite{senousy2021mcua} proposed the MCUa model, which achieved superior effectiveness compared to existing models for breast histology image classification. Huang et al. \cite{huang2022vit} presented ViT-AMCNet, which outperforms existing methods in laryngeal cancer tumor grading. Mahmood  et al. \cite{mahmood2019deep} introduced a nuclei segmentation method using a cGAN trained with synthetic and real data. Li et al. \cite{li2021dual} presented DUALCORENET, a novel end-to-end deep learning framework for mammogram diagnosis. Lou et al. \cite{lou2022pixel} introduced a framework for nuclei instance segmentation in pathology images, which demonstrated performance comparable to fully-supervised approaches despite only having less than 5\% of pixels annotated. Further research is needed to assess these systems' generalizability, scalability, and practical applicability in real-world clinical settings.

\subsection{Oral Cancer Segmentation}


Image segmentation enables specific cancer cell detection at the pixel level, while classification assigns a class label to an entire image. U-Net-based semantic segmentation \cite{ronneberger2015u} assigns a class label to each pixel, which helps to understand the overall structure and composition of the scene. However, the accuracy of oral cancer detection based on image segmentation is relatively lower than image classification. Instance-based segmentation is more informative, as specific pixels of cancer cells can be detected.



In Martino et al. \cite{martino2020deep}, the ORCA dataset was introduced, and four deep learning-based architectures were compared for oral cancer segmentation: SegNet, U-Net, U-Net with VGG-16 encoder, and U-Net with ResNet50 encoder. 
This work uses testing data 
to evaluate the networks' generalization capabilities, resulting in promising segmentation results. However, the study could benefit from optimizing the deep learning model rather than using standard networks. Our study attempts to minimize the network while improving accuracy, offering further advancements in this area.


Dos et al. \cite{dos2021automated} presented a fully convolutional neural network based on the U-Net model for refined segmentation of oral cavity-derived tumor regions in H\&E-stained histological whole slide images. The method uses color features in the HSV color model to identify tissue regions and removes the background and nonrelevant areas. With a total of 10,50 image patches of size 640 × 640 pixels, the method achieved accuracy results up to 97.6\%, specificity up to 98.4\%, and sensitivity up to 92.9\% on the OCDC dataset \cite{dos2021automated}. The study also explored the influence of different color spaces and image-patch sizes. We conduct a comparative evaluation in the results and evaluation section to highlight OCU-Net's advancements and evaluate its generalizability and effectiveness.


In Pennisi et al. \cite{pennisi2022multi}, a modified U-Net architecture called Multi-encoder U-Net is proposed for segmenting 
OSCC in whole slide image samples. The method splits the input image into tiles, with each tile encoded by a separate encoder and merged using a convolutional layer. The resulting merged layers are decoded to obtain the segmented image. The proposed approach achieves promising accuracy results on the ORCA dataset \cite{martino2020deep} but lacks a detailed comparison with other state-of-the-art methods. Additionally, the effectiveness of the proposed method needs to be evaluated on larger datasets to demonstrate its generalization capabilities.

Despite the success of deep learning techniques in various medical image modalities, applying them to H\&E images, particularly for oral cancer detection based on image segmentation, remains limited. Our proposed work is significant and unique, as it employs an optimized U-Net based model and achieves state-of-the-art performance for H\&E image based oral cancer segmentation.






\section{Method}

\subsection{Oral Cancer H\&E Image Datasets}

Our study utilizes two datasets (ORCA \citep{martino2020deep} and  OCDC \citep{dos2021automated}) to train and evaluate the effectiveness of our OCU-Net models for the accurate detection and segmentation of oral cancer. 
The datasets are summarized in Table~\ref{tab:dataset_summary}.

\begin{table*}[]
\caption{Summary of the ORCA and OCDC datasets for image segmentation}
\label{tab:dataset_summary}
\begin{tabular}{|p{1.5cm}|p{7.9cm}|p{1cm}|p{2.5cm}|p{2.5cm}|} \hline
\textbf{Dataset} & \textbf{Description} & \textbf{\# WSI} & \textbf{\# Train Images} & \textbf{\# Test Images} \\
\hline
ORCA \citep{martino2020deep} & H\&E-stained oral cancer annotated (ORCA) dataset with ground-truth data derived from the Cancer Genome Atlas (TCGA) dataset, annotated by two expert pathologists. & 200 & 100 Core images & 100 Core images\\
\hline
OCDC \citep{dos2021automated} & H\&E-stained oral cavity-derived cancer (OCDC) dataset with whole slide images (WSIs) of oral squamous cell carcinoma (OSCC) patients, hand-annotated by a pathologist. & 15 & 840 Image patches & 210 Image patches\\
\hline
\end{tabular}
\end{table*}

\begin{figure*}
    \centering
    \fbox{
    \includegraphics[scale=.3, angle=0]{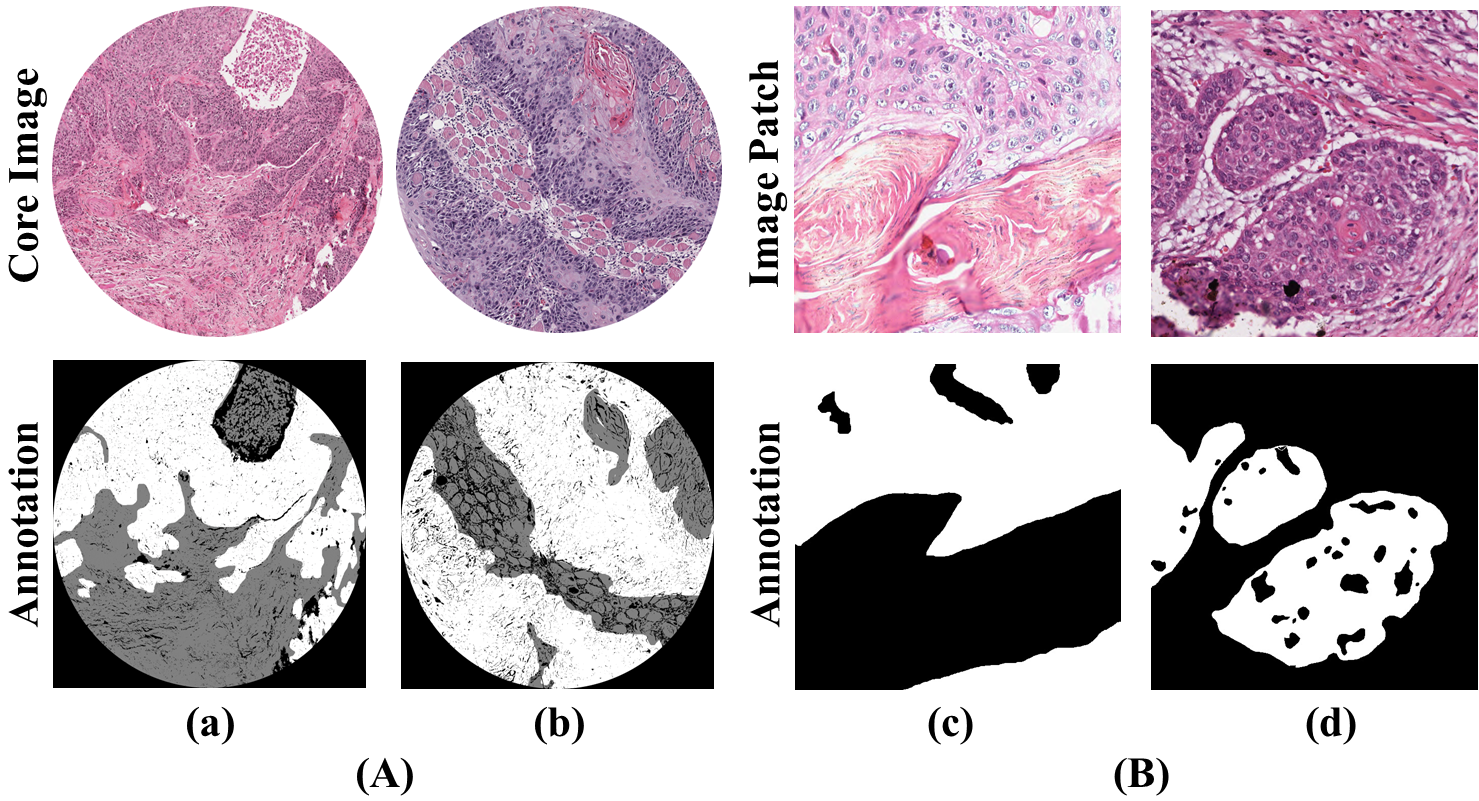}
    }
    \caption{(A) ORCA Dataset image and annotation examples: (a) Training set example. (b) Testing set example. Classes include (White: Carcinoma pixels), (Gray: Non-carcinoma tissue pixels), and (Black: Non-tissue pixels). (B) OCDC Dataset image and annotation examples: (c) Training set example. (d) Testing set example. Classes include: (White: Carcinoma pixels) and (Black: Non-carcinoma tissue pixels).}
    \label{fig:orca-ocdc}
\end{figure*}

The first dataset, ORCA \citep{martino2020deep}, is composed of 200 H\&E-stained WSIs derived from the Cancer Genome Atlas (TCGA) dataset, and manually annotated by two expert pathologists. Each WSI contains one or two cores with a fixed size of 4500$\times$4500 pixels and includes ground-truth data for tumor pixels. The dataset has two subsets, consisting of 100 core images each: a validation set and a test set. Figure \ref{fig:orca-ocdc} (A) displays examples of core images along with their annotations from the validation set (a) and  testing set (b) of the ORCA dataset. Unfortunately, the original training set utilized by Martino et al. \citep{martino2020deep} was not made public. Therefore, we used the validation set as training data for this research, as shown in Table \ref{tab:dataset_summary}.


The second dataset, OCDC \citep{dos2021automated}, consists of 15 H\&E-stained WSIs of human patients diagnosed with OSCC. A total of 1050 image patches of size 640$\times$640 pixels were extracted from the WSIs, and hand-annotated by a pathologist to indicate the presence or absence of tumor regions. The dataset has two subsets: a training set of 840 image patches and a test set of 210 image patches. The OCDC dataset was specifically designed for automated detection and segmentation of OSCC. Figure \ref{fig:orca-ocdc} (B) displays examples of patch images along with their annotations from the training set (c) and testing set (d) of the OCDC dataset \cite{dos2021automated}. 

\begin{figure*}[h!]
    \centering
    \includegraphics[scale=.20, angle=0]
    {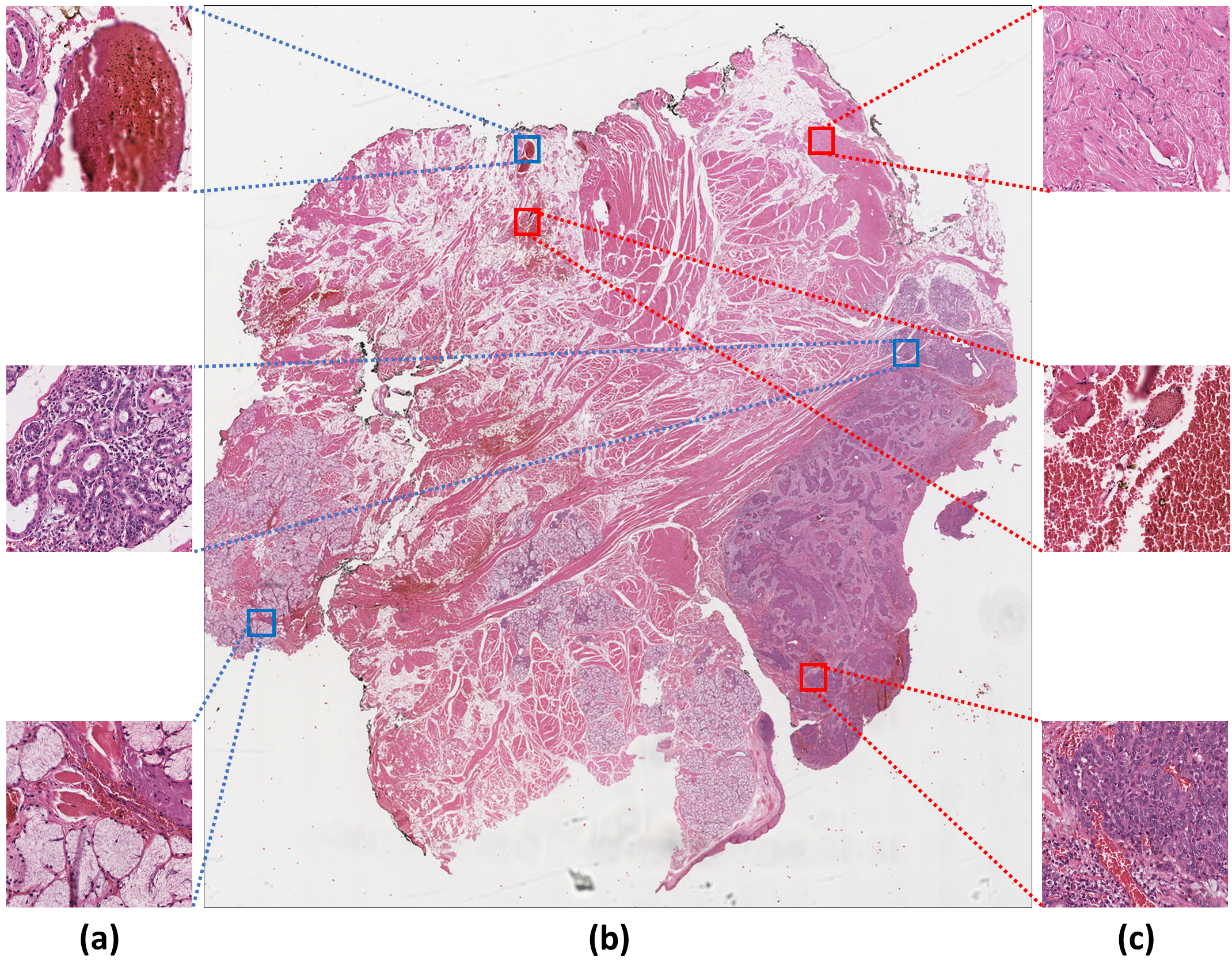}
   \caption{Illustration of H\&E images from the OCDC dataset \cite{dos2021automated}. (a) Blue highlighted regions indicate image patches extracted for the training set. (b) Whole slide image (WSI). (c) Red highlighted regions indicate image patches extracted for the testing set.}
    \label{fig:ocdc-data}
\end{figure*}

Figure~\ref{fig:ocdc-data} further illustrates the WSI and patch H\&E images from the OCDC dataset. Panels (a) and (c) display image patches for the training and testing sets, highlighted in blue and red respectively, that are extracted from a WSI as displayed in Panel (b). The figure provides an overview of the OCDC dataset images used in the study and its division into training and testing sets.



\subsection{Oral Cancer U-Net Segmentation Model}

We introduce OCU-Net and OCU-Net$^m$, two enhanced U-Net-based architectures designed for oral cancer segmentation in H\&E-stained images. The OCU-Net architecture builds upon the original U-Net \citep{ronneberger2015u}. It includes a novel Channel and Spatial Attention Fusion (CSAF) module that refines channel features and captures long-range dependencies to exploit contextual information effectively, thus advancing attention architecture for cancer detection. In addition, OCU-Net includes other powerful components such as the Squeeze-and-Excitation (SE) attention module \citep{hu2018squeeze}, Atrous Spatial Pyramid Pooling (ASPP) module \citep{chen2017deeplab}, residual blocks, and multi-scale fusion. OCU-Net$^m$ integrates the ImageNet pre-trained MobileNet-V2 \citep{singh2019shunt} model as an encoder and feature extractor, utilizing 5.47 million parameters.

\begin{figure*}[h!]
    \centering
    \includegraphics[scale=.15, angle=0]
    {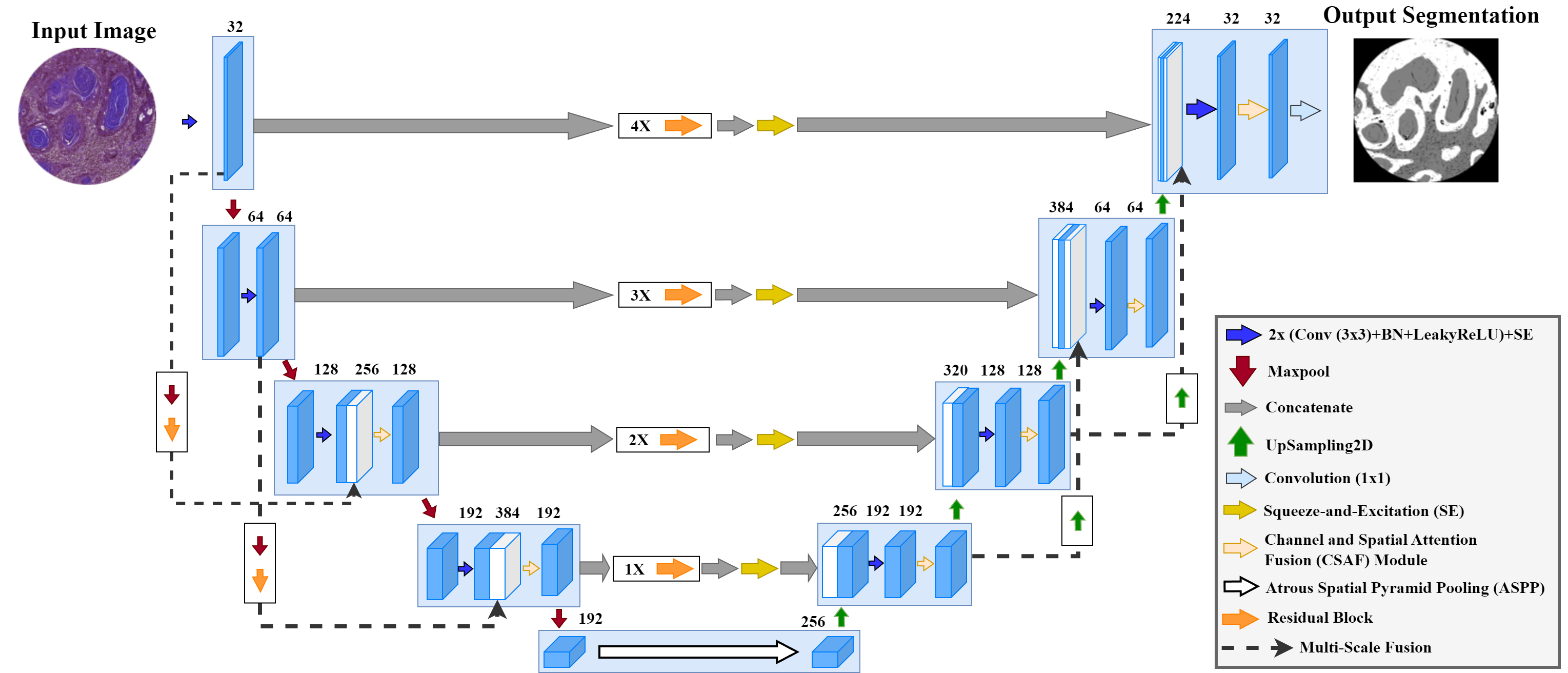}
    \caption{OCU-Net Architecture} 
    \label{fig:ocu-net}
\end{figure*}

The OCU-Net architecture is depicted in Figure~\ref{fig:ocu-net}. In addition to the CSAF module, OCU-Net employs multiple residual blocks on the skip connections to bridge the semantic gap between the encoder and decoder stages, drawing inspiration from \citep{zhou2018unet++, ibtehaz2020multiresunet, yu2021tumor}. The architecture also employs a multi-scale fusion strategy, integrating features from different levels of the network to capture fine-grained details and high-level contextual information. OCU-Net incorporates the Atrous Spatial Pyramid Pooling (ASPP) module as the bottleneck layer of the architecture, which captures multi-scale contextual information through parallel atrous convolutions with varying dilation rates.

The convolution blocks of OCU-Net include squeeze-and-excitation blocks, which provide an attention mechanism for the channel dimension. The combination of these enhancements within OCU-Net 
leads to a robust and efficient architecture for oral cancer segmentation.

Both OCU-Net and OCU-Net$^m$ deliver outstanding segmentation performance, making them practical tools for detecting oral cancer in H\&E-stained images. In the following subsections, we will describe each component of OCU-Net in greater detail.

\subsubsection{Encoder-Decoder Structure}
The OCU-Net architectures consist of an encoder and decoder. The encoder has four levels with convolution blocks followed by SE attention blocks. Residual blocks are applied in the third and fourth levels to the received multi-scale features, which are fused in the CSAF module. To reduce the features spatial dimensions, max-pooling is applied at the end of each of the four levels. The bottleneck level employs the ASPP module to capture multi-scale contextual information. The decoder also has four levels and uses up-sampling for feature map resolution enhancement. It also employs a CSAF module at the end of each level to refine the concatenated contextual information. On the skip connections residual blocks were applied and their number were reduce by one as the depth of the network increases. In the final two levels of the decoder, OCU-Net integrates multi-scale features received from the two previous levels and fuses them using concatenation to be further processed through the Conv2D block and CSAF module. This design enables OCU-Net to effectively capture contextual information and long-range dependencies, resulting in superior segmentation performance compared to traditional U-Net architectures.

The OCU-Net$^m$ utilizes the pre-trained MobileNet-V2 model \citep{singh2019shunt} as its encoder to improve efficiency and reduce computational complexity. MobileNet-V2 model is a popular choice in computer vision tasks and has been widely used due to its effectiveness.
Specifically, it employs depthwise separable convolutions that significantly reduce computational complexity while maintaining accuracy. This enables the OCU-Net$^m$ model to achieve high segmentation accuracy using fewer parameters and less memory. 
The bottleneck, skip connections, and decoder part of the OCU-Net$^m$ is same as OCU-Net.



\subsubsection{Channel and Spatial Attention Fusion Module}\label{CSAFM}
Attention mechanisms play a crucial role in improving image segmentation by refining features, capturing long-range dependencies, and exploiting contextual information. Different attention mechanisms have been proposed, such as the Convolutional Block Attention Module (CBAM) \citep{woo2018cbam} and the Dual Attention Network (DANet) \citep{fu2019dual}, which focus on channel and spatial attention independently. The Squeeze-and-Excitation (SE) block \citep{hu2018squeeze} emphasizes informative features by channel-wise feature recalibration. 

\begin{figure*}[h!]
    \centering
    \includegraphics[scale=0.22, angle=0]
    {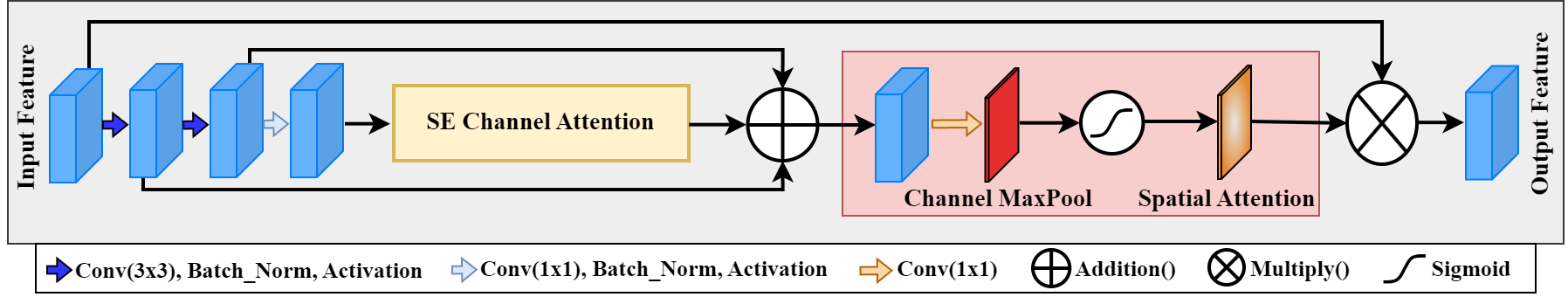}
    \caption{Channel and Spatial Attention Fusion Module (CSAF).}
    \label{fig:CSAFM}
\end{figure*}

We designed and implemented a novel attention module, CSAF, in our network. The structure of CSAF is shown in Figure \ref{fig:CSAFM}. In OCU-Net models, the CSAF module is critical in improving the accuracy and efficacy of oral cancer detection. This novel module integrates both channel and spatial attention while utilizing residual connections for feature fusion, facilitating the effective integration of crucial information across different scales. Additionally, the module incorporates the SE block, simplifying the channel attention process and enhancing the segmentation performance of the model.
The CSAF module is specifically designed to highlight important channel and spatial areas in H\&E images, which are crucial for accurately detecting cancer cells. By exploring contextual information and emphasizing these areas, the module enhances the overall performance of OCU-Net models, improving the segmentation accuracy and facilitating earlier and more accurate oral cancer detection.

OCU-Net includes six CSAF modules, two in the encoder and four in the decoder as seen Figure \ref{fig:ocu-net}. In contrast, OCU-Net$^m$ only includes four CSAF modules, all in the decoder. The following section will provide a detailed discussion of the CSAF module.

As seen in Figure \ref{fig:CSAFM}, the CSAF module processes feature maps through three Conv2D layers, with the first two using a 3x3 kernel and the third a 1x1 kernel, each followed by batch normalization and Leaky ReLU activation. The resulting feature maps are fed into the SE block. The CSAF takes $X\in R^{H×W×C}$ as input, with $H$, $W$, and $C$ representing height, width, and channel dimensions, respectively. Equation \ref{SAB_1} shows the formula for convolution operations in CSAF. In this equation the outputs of the three Conv2D blocks ($F_{1}(X)$, $F_{2}(X)$, and $F_{3}(X)$) are denoted as $F_{k}(X)$, with $C_{k}(X)$ representing the convolution operation, $\mathcal{B}$ representing batch normalization, and $\delta$ representing the Leaky ReLU activation function.

\begin{equation}
F_k(X) = \delta(\mathcal{B}(\mathcal{C}_k(X))),\quad k = 1, 2, 3
\label{SAB_1}
\end{equation}

The feature maps from the third Conv2D block, $F_{3}(X)$, are fed into the SE block which applies global average pooling to achieve a global understanding of each channel and fully connected layer to obtain recalibrated feature maps $SE(F_{3}(X))$. The SE block output is added to the outputs of the first and second Conv2D blocks using addition layer, represented by equation \ref{SAB_2}. The result of this addition is represented by $A(x)$ in this equation. The addition procedure here results in a more integrated manner by employing residual connections.

\begin{equation}
A(X) = F_1(X) + F_2(X) + SE(F_3(X))
\label{SAB_2}
\end{equation}

We also incorporated Spatial Attention (SA) module in the CSAF. In SA, max pooling is applied first across the channel dimension of $A(X)$ to obtain $F_{max}(X)$. Then, a 2D convolution with sigmoid $\sigma$ activation is applied on $F_{max}(X)$ using a kernel size $k_s \in \{5, 7\}$, where $s$ is selected based on the current spatial dimension of the feature map if it is greater than $(128\times128)$ then $s = 7$ otherwise $s = 5$. This yields the spatial attention map $M(X)$, highlighting important spatial regions (Equation \ref{SAB_3}).

\begin{equation}
M(X) = \sigma(\mathcal{C}{k_s}(F{max}(X)))
\label{SAB_3}
\end{equation}

After this, the spatial attention map $M(X)$ is multiplied by the CSAF input, Equation \ref{SAB_4}, retaining original features while focusing on relevant regions. This enhances the model's ability to capture context and long-range dependencies.

\begin{equation}
\begin{aligned}
& Y_{ijc} = M(X){ijc} \cdot X{ijc}\quad  \\& i = 1, \dots, H;\ j = 1, \dots, W;\ c = 1, \dots, C
\end{aligned}
\label{SAB_4}
\end{equation}

Integrating the CSAF module boosts the model's performance by effectively capturing channel interdependencies and selectively concentrating on crucial spatial regions within the image. This channel and spatial information fusion allows the model to understand contextual cues better and adaptively emphasize relevant features, ultimately leading to more accurate and robust image segmentation results.

\subsubsection{Squeeze-and-Excitation Block}

\begin{figure}[h!]
    \centering
    \includegraphics[scale=0.15, angle=0]
    {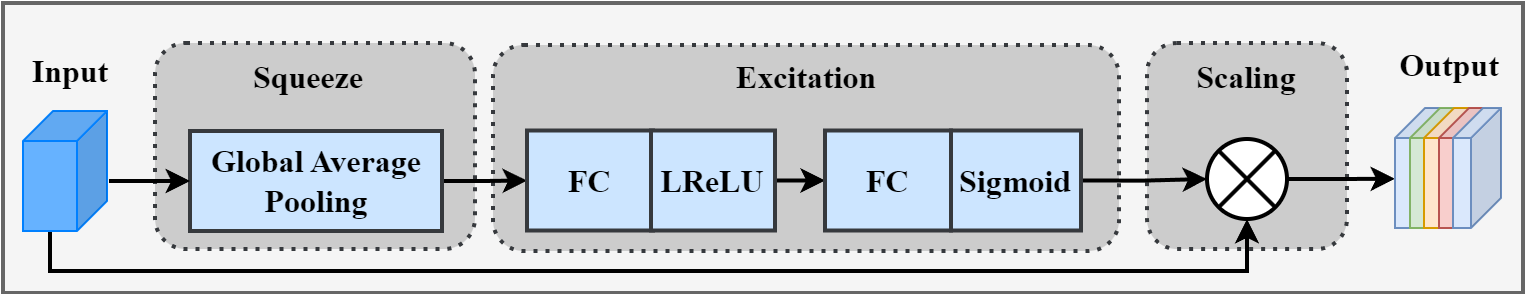}
    \caption{Squeeze-and-Excitation Block (SE).}
    \label{SE}
\end{figure}

The Squeeze-and-Excitation (SE) block was introduced by Hu et al.  \citep{hu2018squeeze} and has been proven to be an effective technique for enhancing the performance of convolutional neural networks (CNNs) in various computer vision tasks. By selectively emphasizing informative features, the SE block enables the network to allocate more resources to important regions of the input data. The SE block consists of three  steps: squeeze, excitation, and scaling. Figure \ref{SE} illustrates the sequence of these three steps. 

\noindent \textbf{Squeeze}: In this step, the SE block summarizes the global spatial information of each channel in the input feature maps by applying global average pooling. This generates a compact descriptor (vector) of size C, where C represents the number of channels. Given an input feature map $X\in R^{H×W×C}$, the squeeze operation computes the descriptor $z\in R^{C}$ as shown in equation \ref{SE_1}. 

\begin{equation}
z_c = \frac{1}{H \times W} \sum_{i=1}^{H} \sum_{j=1}^{W} X_{ijc},\quad c = 1, \dots, C
\label{SE_1}
\end{equation}

\noindent \textbf{Excitation}: This step generates channel-wise weights using the descriptor obtained from the squeezing step. A small fully connected (FC) neural network with a hidden layer is used to learn non-linear interactions between channels. The first FC layer employs a Leaky ReLU activation function, while the second FC layer uses a sigmoid activation function to output channel-wise weights. Let $F_{ex}$ denote the excitation function with hidden layer size $r$, and $W_{1}\in R^{\frac{c}{r}\cdot C}$, $W_{2}\in R^{C \cdot \frac{c}{r}}$ as the weight matrices for the FC layers. The excitation operation computes channel-wise weights $s\in R^{C}$ as shown in equation \ref{SE_2}, where $\delta$ represents the Leaky ReLU activation function and $\sigma$ represents the sigmoid activation function:

\begin{equation}
s = F_{ex}(z, W_1, W_2) = \sigma(W_2 \cdot \delta(W_1 \cdot z))
\label{SE_2}
\end{equation}

The SE block output is obtained by re-scaling the input feature maps using channel-wise weights (Equation \ref{SE_3}). This maintains the spatial dimensions and emphasizes informative channels.

\begin{equation}
\begin{aligned}
& Y_{ijc} = s_c \cdot X_{ijc},\quad \\ & i = 1, \dots, H;\ j = 1, \dots, W;\ c = 1, \dots, C
\end{aligned}
\label{SE_3}
\end{equation}

Our proposed OCU-Net architectures leverage the SE block to capture relevant features while disregarding irrelevant ones, resulting in more accurate and robust segmentation results. Furthermore, including the SE block effectively manages the difficulty of detecting oral cancer due to the significant differences in size, shape, and appearance of cancer. Our experimental results demonstrate the SE block's effectiveness in improving the OCU-Net model's segmentation performance, making it well-suited for accurate oral cancer detection and segmentation.

\subsubsection{Residual Block}
The OCU-Net architecture incorporates residual blocks inspired by \citep{zhou2018unet++, ibtehaz2020multiresunet, yu2021tumor} to enhance the learning process and facilitate information transfer between encoder and decoder stages. These blocks address the vanishing gradient problem and enable efficient information transfer through a series of nonlinear operations. Additionally, they bridge the semantic gap between the encoder and decoder stages by connecting layers with varying receptive fields.

Algorithm \ref{alg:residual_blocks} explains the residual block process. As shown, the function \textit{ApplyResidualBlocks} takes input feature maps ($F_{in}$) and the level number ($L$) to determine the number of residual blocks ($R(L) = 5 - L$) to be applied. It then iteratively calls the \textit{ResidualBlock} function, which consists of parallel 3x3 and 1x1 convolutional layers, batch normalization, and Leaky ReLU activation, followed by a sum operation. At the end of the  residual blocks, a squeeze-and-excitation block for channel attention is applied and output is returned. This process enhances the learning capability and information transfer within the OCU-Net architecture.


In the OCU-Net and OCU-Net$^m$ architectures, the number of residual blocks applied at each level in the decoder is determined by the level number. For example, feature maps from levels one, two, three, and four go through four, three, two, and one residual blocks, respectively. This design choice allows for more efficient information transfer and better feature representation at different scales. In addition, by allocating more residual blocks to shallower levels, the network can better capture and refine low-level features, which are crucial for accurately delineating boundaries and detecting fine-grained details in the segmentation task. Conversely, by decreasing the number of residual blocks in deeper layers, the focus remains on higher-level, abstract representations, which ultimately results in better segmentation performance and greater accuracy in detecting oral cancer.

\begin{algorithm}
\caption{Residual Blocks in OCU-Net Skip-Connections}
\label{alg:residual_blocks}
\begin{algorithmic}[1]
\Function{ApplyResidualBlocks}{$F_{in}, L$}
    \State $F_{out} \gets F_{in}$
    \State $R(L) \gets 5 - L$
    \For{$i \gets 1$ to $R(L)$}
        \State $F_{out} \gets \Call{ResidualBlock}{F_{out}}$
    \EndFor
    \State \Return  $\Call{SE}{F_{out}}$
\EndFunction

\vspace{1em}

\Function{ResidualBlock}{$F_{in}$}
    \State $F_{res} \gets F_{in}$
    \State $F_{3x3} \gets \Call{Conv3x3}{F_{in}}$, $F_{1x1} \gets \Call{Conv1x1}{F_{in}}$
    \State $F_{3x3} \gets \Call{BN}{F_{3x3}}$, $F_{1x1} \gets \Call{BN}{F_{1x1}}$
    \State $F_{3x3} \gets \Call{LReLU}{F_{3x3}}$, $F_{1x1} \gets \Call{LReLU}{F_{1x1}}$
    \State $F_{sum} \gets F_{3x3} + F_{1x1}$
    \State \Return $F_{sum}$
\EndFunction
\end{algorithmic}
\end{algorithm}

\subsubsection{Multi-Scale Fusion}
The OCU-Net architecture effectively captures contextual information and handles objects of different sizes and shapes through multi-scale fusion. Specifically, the third and fourth levels in the encoder and levels eight and nine in the decoder receive inputs from their two preceding levels. Figure \ref{fig:ocu-net} illustrates the multi-scale fusion in OCU-Net with dashed lines. On the encoder side, low-level features are down-sampled and refined through a residual block before concatenation. In contrast, high-level features are up-sampled and concatenated on the decoder side since they require less refinement. The integration of features from different resolutions through multi-scale fusion enhances segmentation accuracy and precision. Notably, OCU-Net$^m$ utilizes multi-scale fusion only at the decoder level.

Several previous studies have also explored the use of multi-scale fusion in segmentation tasks. For instance, Zhou et al. \cite{zhou2018unet++} proposed the U-Net++ architecture, which incorporates nested and dense skip connections to improve the information flow between different levels. Similarly, Ibtehaz et al. \cite{ibtehaz2020multiresunet} introduced the MultiResUNet architecture, which employs a multi-resolution fusion mechanism to merge features from multiple resolutions, leading to enhanced performance in medical image segmentation tasks.

In both versions of OCU-Net, multi-scale fusion offers various advantages. It captures both local and global contextual information and handles objects of different sizes and shapes. This approach helps the networks to learn more distinct features, improve generalization, and achieve better segmentation performance than traditional U-Net architectures.

\subsubsection{Atrous Spatial Pyramid Pooling}
\begin{figure}[h!]
    \centering
    \includegraphics[scale=0.185, angle=0]
    {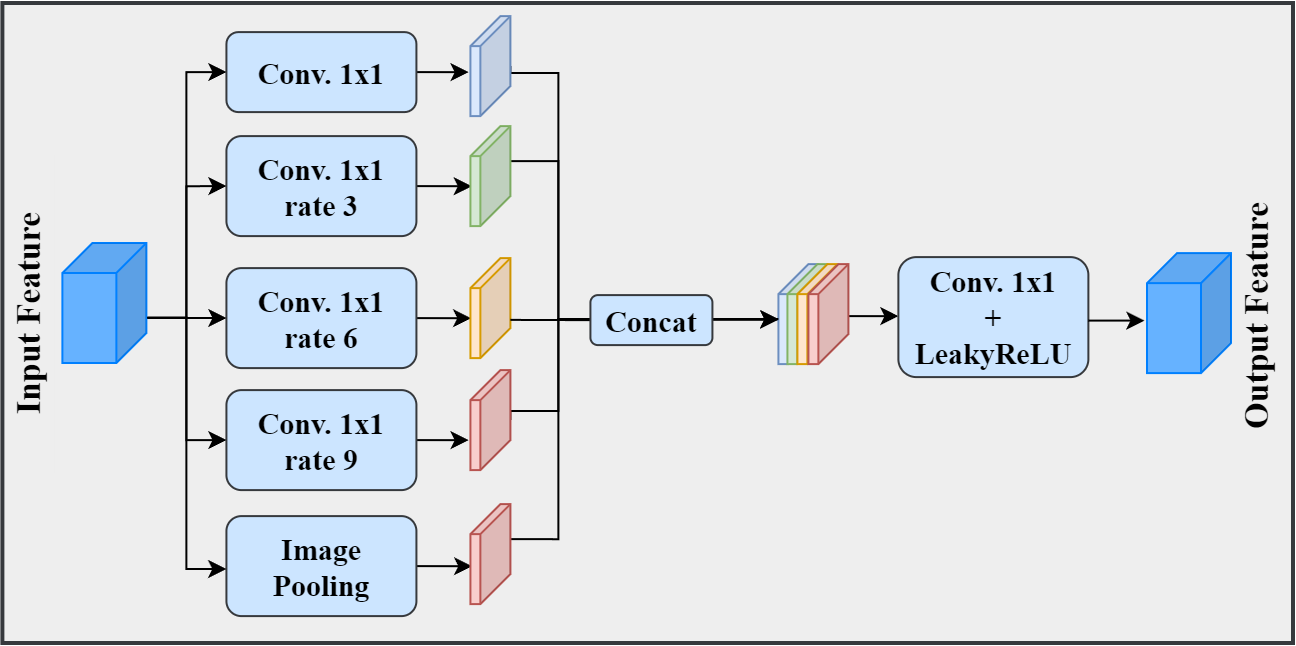}
    \caption{Atrous Spatial Pyramid Pooling (ASPP).}
    \label{aspp}
\end{figure}

The ASPP module is a crucial component of both versions of OCU-Net architecture, acting as the bottleneck layer to capture multi-scale contextual information. Introduced by Chen et al. \cite{chen2017deeplab}, the module employs parallel atrous convolutions with varying dilation rates, as seen in Figure \ref{aspp}, to effectively aggregate features across different spatial scales. By incorporating the ASPP module, OCU-Net and OCU-Net$^m$ can robustly handle objects of different sizes and shapes and capture finer details and global context. This is particularly useful in the challenging domain of oral cancer detection, where cancer cells exhibit high variability in size, shape, and appearance. In addition, the ASPP module allows the networks to attend to both local and global contexts, leading to more accurate and precise segmentation results for oral cancer detection.

%

\subsubsection{Loss Function}



We evaluated our proposed architecture using two loss functions, depending on the target segmentation task: Binary vs. Multi-class. For the ORCA dataset, the multi-class loss function, Categorical Cross Entropy (CCE), was used for training, while the binary segmentation task on the OCDC dataset utilized the hybrid loss function comprising Weighted Binary Cross-Entropy (WBCE) \citep{pihur2007weighted} and Dice Loss \citep{dice1945measures}. 

The CCE loss measures the difference between the predicted probability distribution and the true probability distribution for the pixel-wise multi-class classification task on the ORCA dataset. The loss is calculated as the negative log-likelihood of the correct class and is formulated in Equation \ref{CCE}, where $C$ represents the number of classes, $y_{true, i}$ represents the ground truth probability for class $i$, and $y_{pred, i}$ signifies the predicted probability for class $i$.

\begin{equation}
L_{CCE} = -\sum_{i=1}^{C} y_{true, i} \log(y_{pred, i})
\label{CCE}
\end{equation}


For binary segmentation, we use a hybrid loss function, $L = L_{W} + L_{D}$, which combines two components:  Weighted Binary Cross-Entropy (WBCE) and Dice Loss. The WBCE component addresses the common class imbalance issue in medical image segmentation problems. It assigns different weights ($w_i$) to each class to encourage the model to learn the under-represented class better. The WBCE loss is calculated using Equation \ref{WBCE}, where $N$ is the total number of pixels, $y_{true, i}$ is the ground truth label for pixel $i$, $y_{pred, i}$ is the predicted probability for pixel $i$, and $w_i$ is the weight assigned to each pixel based on its class imbalance.


\begin{equation}
\begin{aligned}
L_{W} = & -\frac{1}{N} \sum_{i=1}^{N} w_i \cdot y_{true, i} \cdot \log(y_{pred, i}) \\ & + (1-y_{true, i}) \cdot \log(1-y_{pred, i})
\end{aligned}
\label{WBCE}
\end{equation}
where $N$ represents the total number of pixels.
$y_{true, i}$ denotes the ground truth label for pixel $i$.
$y_{pred, i}$ indicates the predicted probability for pixel $i$.
$w_i$ is the weight assigned to each pixel based on its class imbalance.


The Dice Loss \citep{dice1945measures} measures the overlap between the predicted and ground truth segmentation, taking into account both true positives and false positives/negatives when computing the error. This sensitivity to segmentation quality allows the model to generate more accurate results. The Dice Loss is calculated using Equation \ref{DiceL}, where $y_{true, i}$ and $y_{pred, i}$ represent the ground truth label and predicted probability for pixel $i$, respectively.

\begin{equation}
L_{D} = 1 - \frac{2\sum_{i=1}^{N} y_{true, i} \cdot y_{pred, i}}{\sum_{i=1}^{N} y_{true, i} + \sum_{i=1}^{N} y_{pred, i}}
\label{DiceL}
\end{equation}

The proposed hybrid loss function combines the strengths of WBCE and Dice Loss to address challenges in oral cancer segmentation tasks. The WBCE component addresses class imbalance, while Dice Loss promotes segmentation quality by emphasizing precise predictions. Additionally, the loss function can be fine-tuned using a hyperparameter $\alpha$ to adjust the relative weight of the two components, allowing for customization to specific datasets or problems. The hybrid loss function is defined in Equation \ref{HybridLoss}, where $L_{W}$ and $L_{D}$ represent the Weighted Binary Cross-Entropy loss and the Dice loss, respectively. The hyperparameter $\alpha$ controls the relative weight of the two loss components, allowing for better customization to specific problems or datasets.


\begin{equation}
L_{H} = \alpha \cdot L_{W} + (1-\alpha) \cdot L_{D}
\label{HybridLoss}
\end{equation}

\section{Experiments and Results}

\subsection{Measures}
To assess the performance of our models in oral cancer segmentation, we employed the following evaluation metrics: Dice Similarity Coefficient (DSC), Intersection over Union (IoU), Mean Intersection over Union (mIoU), Sensitivity, Specificity, Precision, and Accuracy. DSC and IoU are especially significant because they gauge the degree of overlap between the predicted segmentation and the actual ground truth masks, which is a reliable measure of segmentation accuracy. All metrics are detailed below:


\begin{equation}
Dice Score = \frac{2TP}{2TP+FP+FN}
\label{Dice}
\end{equation}


\begin{equation}
IoU = \frac{\text{Area of Overlap}}{\text{Area of Union}} = \frac{TP}{TP+FP+FN}
\label{iou}
\end{equation}

\begin{equation}
mIoU = \frac{1}{N} \sum_{i=1}^{N} \frac{TP_i}{TP_i + FP_i + FN_i}
\label{miou}
\end{equation}


\begin{equation}
Sensitivity = \frac{TP}{TP + FN}
\label{Sensitivity}
\end{equation}


\begin{equation}
Specificity = \frac{TN}{TN + FP}
\label{Specificity}
\end{equation}


\begin{equation}
Precision = \frac{TP}{TP + FP}
\label{precision}
\end{equation}


\begin{equation}
Accuracy = \frac{TP + TN}{TP + FP + TN+ FN}
\label{Accuracy}
\end{equation}

The TP, TN, FP, and FN represent true positive, true negative, false positive, and false negative, respectively.

\subsection{Experimental settings}
In this study, we address segmentation problems for both the ORCA dataset, a multiclass problem, and the OCDC, a binary segmentation problem. 
During the training phase, we optimize the network using the Adam optimization algorithm \citep{kingma2014adam} with a learning rate of ($3 x 10^{-4}$). The proposed methods and their experimental evaluation were implemented using the TensorFlow framework on a GPU cluster instance equipped with Nvidia RTX A6000 graphic card. To optimize the training process, we adjusted the batch size according to the image size. Specifically, for images sized 512x512, we used a batch size of eight, while for the larger OCDC images sized 640x640, a batch size of four was used due to hardware limitations.

To address the limited size of our training set and mitigate overfitting, we utilized data augmentation techniques, including horizontal and vertical flipping, image blurring, and image sharpening using Gaussian filters. In addition, we applied two training strategies: early stopping and ReduceLROnPlateau. Early stopping prevents overfitting by monitoring the validation metric,  while ReduceLROnPlateau reduces the learning rate if the validation Dice index metric does not improve for a specified number of epochs.

\subsection{U-Net Architecture Performance Comparison}

Below are brief descriptions of the baseline algorithms.

\begin{itemize}[noitemsep,topsep=0pt,leftmargin=*]
    \item U-Net \citep{ronneberger2015u}: A CNN architecture specifically designed for medical image segmentation, comprising an encoder and a decoder. The encoder extracts features from the input image, while the decoder produces a segmentation map by upsampling the extracted features.
\begin{itemize}[noitemsep,topsep=0pt,leftmargin=*]
        \item With the ORCA dataset, various algorithms have been developed for oral cancer detection, including U-Net for oral cancer detection, U-Net + RGB, and Multi-Encoder U-Net \citep{pennisi2022multi}. U-Net + ResNet-50 has also been developed for oral cancer detection using the ORCA dataset \citep{martino2020deep}.
        \item With the OCDC dataset, U-Net has been used for oral cancer detection with three color modalities: Lab, HSV, and RGB \citep{dos2021automated}.
        \end{itemize}
    \item U-Net variations: Several U-Net variations were evaluated on the ORCA and OCDC datasets, including U-Net++, Att\_U-Net, and U$^2$-Net. The models were implemented using the code available here \cite{keras-unet-collection}.
\end{itemize}

\begin{table}[h]
\caption{Comparison of different models based on the number of parameters in millions (M) and size in megabytes (MB).}
\begin{tabular}{l|l|l}
\hline
Method                                                 & Parameters (M) & Size (MB) \\ \hline
U-Net++ \citep{zhou2018unet++}        & 7.85           & 29.9      \\ \hline
Att\_U-Net \citep{oktay2018attention} & 7.98           & 30.46     \\ \hline
U$^2$-Net \citep{qin2020u2}           & 12.27          & 46.8      \\ \hline
OCU-Net (ours)                                         & 11.39          & 43.46     \\ \hline
OCU-Net$^m$ (ours)                                     & {\bf5.47}           & {\bf20.88}     \\ \hline
\end{tabular}
\label{compare_models}
\end{table}

Table \ref{compare_models} presents a comparison of the models, revealing that OCU-Net$^m$ has the smallest model size at 20.88 MB and the fewest number of parameters, with 5.47 million. OCU-Net has a model size of 43.46 MB and 11.39 million parameters, while U-Net++ and Att\_U-Net have similar sizes of ~30 MB and 7.85 million and 7.98 million parameters, respectively. U$^2$-Net is the largest model with a size of 46.8 MB and 12.27 million parameters. The comparison shows that both versions of OCU-Net have a reasonable model size and parameter count, making them efficient and practical for oral cancer segmentation tasks.

\subsection{Ablation Study}


Ablation studies are crucial in evaluating model performance, allowing researchers to assess the importance and contribution of individual features. Two ablation studies were conducted on the ORCA dataset. Table~\ref{ablation} shows the results of the first study, which compares different model variations and demonstrates the incremental improvements gained by incorporating additional features. Table~\ref{ablation2} presents the results of the second study, which compares the performance of different attention methods on the ORCA dataset. It is worth noting that the experiments were conducted using identical settings and evaluated the models using three performance metrics: Accuracy (Acc \%), Dice Similarity Coefficient (Dice \%), and Mean Intersection over Union (mIoU \%).

\begin{table*}[h]
\caption{Results of Ablation study on the ORCA dataset: Accuracy (Acc \%), Dice Similarity Coefficient (Dice \%), and Mean Intersection over Union(mIoU \%).}
\begin{tabular}{l|l|l|l}
\hline
Method & Acc & Dice & mIoU \\ \hline
Baseline U-Net & 81.17 & 65.12 & 54.52 \\ \hline
Model-1 (SE) & 85.7 & 75.81 & 64.733 \\ \hline
Model-2 (SE+Residual) & 88.31 & 81.7 & 71.49 \\ \hline
Model-3 (SE+Residual+MultiScale) & 89.45 & 83.28 & 73.65 \\ \hline
Model-4 (SE+Residual+MultiScale+ASPP) & 89.44 & 83.55 & 73.7 \\ \hline
Model-5 (OCU-Net) & {\bf 90.02} & {\bf 84.36} & {\bf 75.04} \\ \hline
\end{tabular}
\label{ablation}
\end{table*}

\begin{table*}[]
\caption{Comparison of CSAF module and other attention modules on the ORCA: Carcinoma Class's Accuracy (Acc \%), Dice Similarity Coefficient (Dice \%), and Mean Intersection over Union (mIoU \%).}
\begin{tabular}{l|l|l|l}
\hline
Method                 & Acc   & Dice  & mIoU  \\ \hline
Baseline (U-Net)  \cite{ronneberger2015u}     & 80.22 & 73.41 & 60.57 \\ \hline
Baseline + SE   \cite{hu2018squeeze}       & 82.68 & 78.86 & 66.56 \\ \hline
Baseline + DANet  \cite{fu2019dual}     & 83.43 & 77.9  & 64.96 \\ \hline
Baseline + CBAM \cite{woo2018cbam}        & 84.47 & 77.84 & 65.76 \\ \hline
Baseline + CSAF (ours) & {\bf 84.58} & {\bf 79.83} & {\bf 67.72} \\ \hline
\end{tabular}
\label{ablation2}
\end{table*}

Different variations of the U-Net architecture were compared in an ablation study on the ORCA dataset. The baseline U-Net model achieved 81.17\% accuracy, 65.12\% Dice, and 54.52\% mIoU. By adding Squeeze-and-Excitation (SE) blocks, Model-1 achieved 85.7\% accuracy, 75.81\% Dice, and 64.73\% mIoU. Model-2 improved upon Model-1 by adding residual connections, achieving 88.31\% accuracy, 81.7\% Dice, and 71.49\% mIoU. Model-3 incorporated multi-scale input, resulting in 89.45\% accuracy, 83.28\% Dice, and 73.65\% mIoU. Model-4 added Atrous Spatial Pyramid Pooling (ASPP) to Model-3, achieving 89.44\% accuracy, 83.55\% Dice, and 73.7\% mIoU. Finally, Model-5 integrated the Channel Spatial Attention Fusion (CSAF) module into Model-4, leading to an accuracy of 90.02\%, Dice of 84.36\%, and mIoU of 75.04\%. Based on these results, the best-performing model, Model-5, referred to as OCU-Net, was selected for subsequent evaluation.

The ablation study in Table \ref{ablation2} compares the performance of different attention methods on the ORCA dataset, focusing on the Carcinoma Class. The evaluation metrics used are Accuracy (Acc \%), Dice Similarity Coefficient (Dice \%), and Mean Intersection over Union (mIoU \%). The baseline U-Net model achieved an accuracy of 80.22\%, a Dice score of 73.41\%, and a mIoU of 60.57\%. When integrating different attention mechanisms, the performance improved: Baseline + SE \cite{hu2018squeeze} achieved Acc 82.68\%, Dice 78.86\%, mIoU 66.56\%; Baseline + DANet \cite{fu2019dual} had Acc 83.43\%, Dice 77.9\%, mIoU 64.96\%; and Baseline + CBAM \cite{woo2018cbam} reached Acc 84.47\%, Dice 77.84\%, mIoU 65.76\%. The CSAF module, specifically designed for the OCU-Net, achieved the best results among all the methods, with an accuracy of 84.58\%, a Dice score of 79.83\%, and mIoU of 67.72\%. These results demonstrate the strength of the CSAF module in enhancing the model's performance compared to other attention mechanisms. 

\subsection{Experiment with ORCA Dataset}

We evaluated the performance of our proposed models on the ORCA dataset and compared it with several state-of-the-art methods, including U-Net, U-Net + RGB, Multi-Encoder U-Net, U-Net++, U$^2$-Net, and Att\_U-Net. The evaluation metrics used were Accuracy (Acc \%), Dice Similarity Coefficient (Dice \%), and Mean Intersection over Union (mIoU \%).

Table~\ref{ORCA_average} presents the comparison of average results for the different segmentation methods. Our proposed model, OCU-Net, outperforms the baseline methods with an Accuracy of 90.02\%, a Dice Similarity Coefficient of 84.36\%, and a Mean Intersection over Union of 75.04\%. Moreover, OCU-Net$^m$ achieves the highest performance with an Accuracy of 90.98\%, a Dice Similarity Coefficient of 86.14\%, and a Mean Intersection over Union of 77.10\%. These results demonstrate that our proposed method achieves superior segmentation performance compared to existing state-of-the-art methods on the ORCA dataset.

\begin{table*}[h]
\caption{Average results performance on the ORCA dataset, compared with U-net, U-Net++, U$^2$-Net, and Att\_U-Net. Using Accuracy (Acc \%), Dice Similarity Coefficient (Dice \%), Mean Intersection over Union (mIoU \%).}
\begin{tabular}{c|c|c|c}
\hline
Method   & Acc & Dice  & mIoU  \\ \hline
U-Net  \citep{pennisi2022multi}             & 67       & 68    & 58  \\ \hline
U-Net + RGB   \citep{pennisi2022multi}      & 80       & 80    & 66  \\ \hline
Multi-Encoder U-Net \citep{pennisi2022multi} & 82       & 82    & 72    \\ \hline \hline
U-Net++ \citep{zhou2018unet++} & 86.61   & 78.07    & 66.23   \\ \hline
U$^2$-Net \citep{qin2020u2} & 88.21  & 81.88    & 71.07   \\ \hline
Att\_U-Net \citep{oktay2018attention} & 90.29 & 84.28    & 74.91   \\ \hline \hline
OCU-Net (ours) & 90.02  & 84.36    & 75.04  \\ \hline
OCU-Net$^m$ (ours) & \textbf{90.98}   & \textbf{86.14} & \textbf{77.10}  \\ \hline
\end{tabular}
\label{ORCA_average}
\end{table*}

For the second experiment using the ORCA dataset, we evaluated the proposed model's class-wise performance across Non-Tissue, Tissue Non-Carcinoma, and Carcinoma classes using Accuracy, Dice Similarity Coefficient, and Mean Intersection over Union metrics. Table~\ref{ORCA_classes} presents the per-class segmentation results for various methods, including our OCU-Net and OCU-Net$^m$ architectures, benchmark algorithms such as U-Net + ResNet50 \citep{martino2020deep}, and three models from \citep{pennisi2022multi}: U-Net, U-Net + RGB, and Multi-Encoder U-Net. We also compared cancer detection using popular U-Net architectures, including U-Net++ \citep{zhou2018unet++}, Att\_U-Net \citep{oktay2018attention}, and U$^2$-Net \citep{qin2020u2}, aiming to demonstrate the effectiveness of our OCU-Net and OCU-Net$^m$ models in relation to existing methods.

The results show that both OCU-Net models outperform all other methods in terms of all evaluation metrics for all three classes. Specifically, OCU-Net$^m$ achieved an accuracy of 97.12\%, a Dice Similarity Coefficient of 95.31\%, and a mean intersection over the union of 91.28\% for the Non-Tissue class, 86.96\%, 79.13\%, and 66.51\% for the Tissue Non-Carcinoma class, and 88.87\%, 83.96\%, and 73.52\% for the Carcinoma class. Overall, the proposed methods demonstrate superior segmentation performance on all three classes of the ORCA dataset compared to existing methods, with a significant improvement in detecting the Carcinoma class.

\begin{table*}[h]
\caption{Per class results performance on the ORCA dataset, compared with U-net, U-Net++, U$^2$-Net, and Att\_U-Net. Using Accuracy (Acc \%), Dice Similarity Coefficient (Dice \%), and Mean Intersection over Union (mIoU \%). Not reported (-).}
\begin{tabular}{c|c|c|c|c|c|c|c|c|c}
\hline
\multirow{2}{*}{Method} & \multicolumn{3}{c|}{\bf Non-Tissue} & \multicolumn{3}{c|}{\bf Tissue Non-Carcinoma} & \multicolumn{3}{c}{\bf Carinoma} \\ \cline{2-10} 
 & Acc & Dice & IoU & Acc & Dice & IoU & Acc & Dice & IoU \\ \hline
U-Net + ResNet-50 \citep{martino2020deep} & - & - & 85 & - & - & 59 & - & - & 56 \\ \hline
U-Net \citep{pennisi2022multi} & - & - & 79 & - & - & 52 & - & - & 54 \\ \hline
U-Net + RGB \citep{pennisi2022multi} & - & - & 91 & - & - & 54 & - & - & 53 \\ \hline
Multi-Encoder U-Net \citep{pennisi2022multi}& - & - & 96 & - & - & 63 & - & - & 58 \\ \hline \hline
 U-Net++ \citep{zhou2018unet++} & 92.69 & 87.21 & 77.99 & 81.85 & 69.09 & 54.44 & 85.29 & 77.9 & 66.24 \\ \hline
 U$^2$-Net \citep{qin2020u2} & 94.56 & 94.49 & 83.05 & 84.16 & 73.89 & 60.19 & 85.9 & 81.26 & 69.96 \\ \hline 
 Att\_U-Net \citep{oktay2018attention}& 96.76 & 94.87 & 90.49 & 85.96 & 76.76 & 63.5 & 88.15 & 81.21 & 70.75 \\ \hline \hline
OCU-Net (ours) & 96.75 & 94.54 & 89.81 & 85.66 & 75.8 & 62.91 & 87.66 & 82.74 & 72.4 \\ \hline
OCU-Net$^m$ (ours) & \textbf{97.12} & \textbf{95.31} & \textbf{91.28} & \textbf{86.96} & \textbf{79.13} & \textbf{66.51} & \textbf{88.87} & \textbf{83.96} & \textbf{73.53} \\ \hline
\end{tabular}
\label{ORCA_classes}
\end{table*}

\subsection{Experiment with OCDC Dataset}
In our study, we also conducted an experiment utilizing the OCDC dataset to evaluate the performance of the proposed OCU-Net architectures. In addition, we compared our architectures with several benchmark algorithms, such as the U-Net based oral cancer segmentation on the OCDC dataset, as presented by Dos et al. \cite{dos2021automated}. Moreover, we investigated cancer detection using popular U-Net architectures, including U-Net++ \citep{zhou2018unet++}, Att\_U-Net \citep{oktay2018attention}, and U$^2$-Net \citep{qin2020u2}.

We present the average results of different models on the OCDC dataset using various performance metrics. Table~\ref{OCDC_results} shows results for image patches with a size of 640x640, while Table~\ref{OCDC_results_512} shows results for image patches with a size of 512x512. In terms of the best performance for the 640x640 image patch size Table~\ref{OCDC_results}, the OCU-Net$^m$ model outperforms the others. It has the highest Accuracy of 98.07\%, Precision of 93.55\%, F1/Dice of 93.49\%, intersection over union (IoU) of 87.78\%, Sensitivity of 93.44\%, and Specificity of 98.88\%. For the 512x512 image patch size Table~\ref{OCDC_results_512}, while the U$^2$-Net model has the highest Accuracy of 98.38\%, the OCU-Net$^m$ model achieves the best results in Precision of 94.62\%, F1/Dice of 94.09\%, and IoU of 88.84\%, Sensitivity of 93.57\%, and Specificity of 99.06\% in this case as well.

\begin{table*}[h]
\caption{Average results performance for the test image patches on the OCDC dataset with an image patch size of 640 x 640. Using Accuracy (Acc \%), Precision (Prec. \%), Dice Similarity Coefficient (Dice \%), intersection over union(IoU \%), Sensitivity (Sens. \%), and Specificity (Spec. \%).}
\begin{tabular}{c|c|c|c|c|c|c|c}
\hline
Model & Color Model & Acc. & Prec. & F1/Dice & IoU & Sens. & Spec. \\ \hline

                                & L*a*b* & 97.6 & 91.1 & 92.0 & 85.2 & 92.9 & 98.4 \\ \cline{2-8}
U-Net \citep{dos2021automated} & HSV & 97.1 & 89.2 & 90.3 & 82.4 & 91.5 & 98.0 \\ \cline{2-8}
                               & RGB & 97.2 & 91.0 & 90.8 & 83.1 & 90.5 & 98.4 \\ \hline \hline
U-Net++ \citep{zhou2018unet++} & RGB & 97.30 & 90.77 & 90.90 & 83.32 & 91.03 & 98.39 \\ \hline
Att\_U-Net \citep{oktay2018attention} & RGB & 97.45 & 92.17 & 91.32 & 84.03 & 90.48 & 98.66 \\ \hline
U$^2$-Net \citep{qin2020u2} & RGB & 97.77 & 92.49 & 92.49 & 86.02 & 92.49 & 98.69 \\ \hline \hline
OCU-Net (ours) & RGB & 97.91 & 93.33 & 92.92 & 86.77 & 92.51 & 98.85 \\ \hline 
OCU-Net$^m$ (ours) & \textbf{RGB} & \textbf{98.07} & \textbf{93.55} & \textbf{93.49} & \textbf{87.78} & \textbf{93.44} & \textbf{98.88} \\ \hline

\end{tabular}
\label{OCDC_results}
\end{table*}

The U-Net study conducted by \citep{dos2021automated} examined deep learning applications in medical imaging, utilizing the OCDC dataset and three color models:  {L*a*b}, HSV, and RGB (see Table~\ref{OCDC_results}). Although the study initially suggested that the  {L*a*b}  model outperformed the others, our experimental results revealed that the RGB model was superior to both Lab and HSV in terms of performance.

The {L*a*b} color model divides color information into three channels: lightness (L) and two color channels ({a*} and {b*}). This model is renowned for its accurate representation of color information and is frequently employed in medical imaging for image analysis. In contrast, the HSV model separates color information into hue, saturation, and value, making it a popular choice in medical imaging for object segmentation based on color attributes. Lastly, the RGB model segregates color information into primary red, green, and blue channels. This model is commonly used in medical imaging for visual analysis and display, as it can represent a wide range of colors and is easily viewed on most computer monitors. The selection of a color model in medical imaging is contingent upon the specific application and image characteristics, as each model possesses unique strengths and weaknesses that can influence the accuracy and efficacy of image analysis techniques.

In summary, the OCU-Net$^m$ model consistently performs well across both image patch sizes and is generally the top performer. These findings suggest that the OCU-Net$^m$ architecture with the MobileNetV2 backbone encoder is highly effective for automated oral cancer screening and holds the potential to enhance the accuracy and efficiency of the screening process.

\begin{table*}[h]
\caption{Average results performance for the test image patches on the OCDC dataset with an image patch size of 512 x 512. Using Accuracy (Acc \%), Precision (Prec. \%), Dice Similarity Coefficient (Dice \%), intersection over union(IoU \%), Sensitivity (Sens. \%), and Specificity (Spec. \%).}
\begin{tabular}{c|c|c|c|c|c|c|c}
\hline
Model & Color Model & Acc. & Prec. & F1/Dice & IoU & Sens. & Spec. \\ \hline

                                & L*a*b* & 97.6 & 91.1 & 92.0 & 85.2 & 92.9 & 98.4 \\ \cline{2-8}
U-Net (640*640) \citep{dos2021automated} & HSV & 97.1 & 89.2 & 90.3 & 82.4 & 91.5 & 98.0 \\ \cline{2-8}
                               & RGB & 97.2 & 91.0 & 90.8 & 83.1 & 90.5 & 98.4 \\ \hline \hline
U-Net++ \citep{zhou2018unet++} & RGB & 97.92 & 89.96 & 91.39 & 84.15 & 92.87 & 98.17 \\ \hline
Att\_U-Net \citep{oktay2018attention} & RGB & 98.03 & 91.32 & 91.64 & 84.58 & 91.97 & 98.46 \\ \hline
U$^2$-Net \citep{qin2020u2} & RGB & \textbf{98.38} & 93.42 & 92.71 & 86.41 & 92.02 & 98.86 \\ \hline \hline
OCU-Net (ours) & RGB & 98.07 & 93.59 & 93.57 & 87.92 & 93.55 & 98.87 \\ \hline
OCU-Net$^m$ (ours) & \textbf{RGB} & 98.24 & \textbf{94.62} & \textbf{94.09} & \textbf{88.84} & \textbf{93.57} & \textbf{99.06} \\ \hline

\end{tabular}
\label{OCDC_results_512}
\end{table*}


\begin{figure*}[h!]
    \centering
    \includegraphics[scale=.4, angle=0]
    {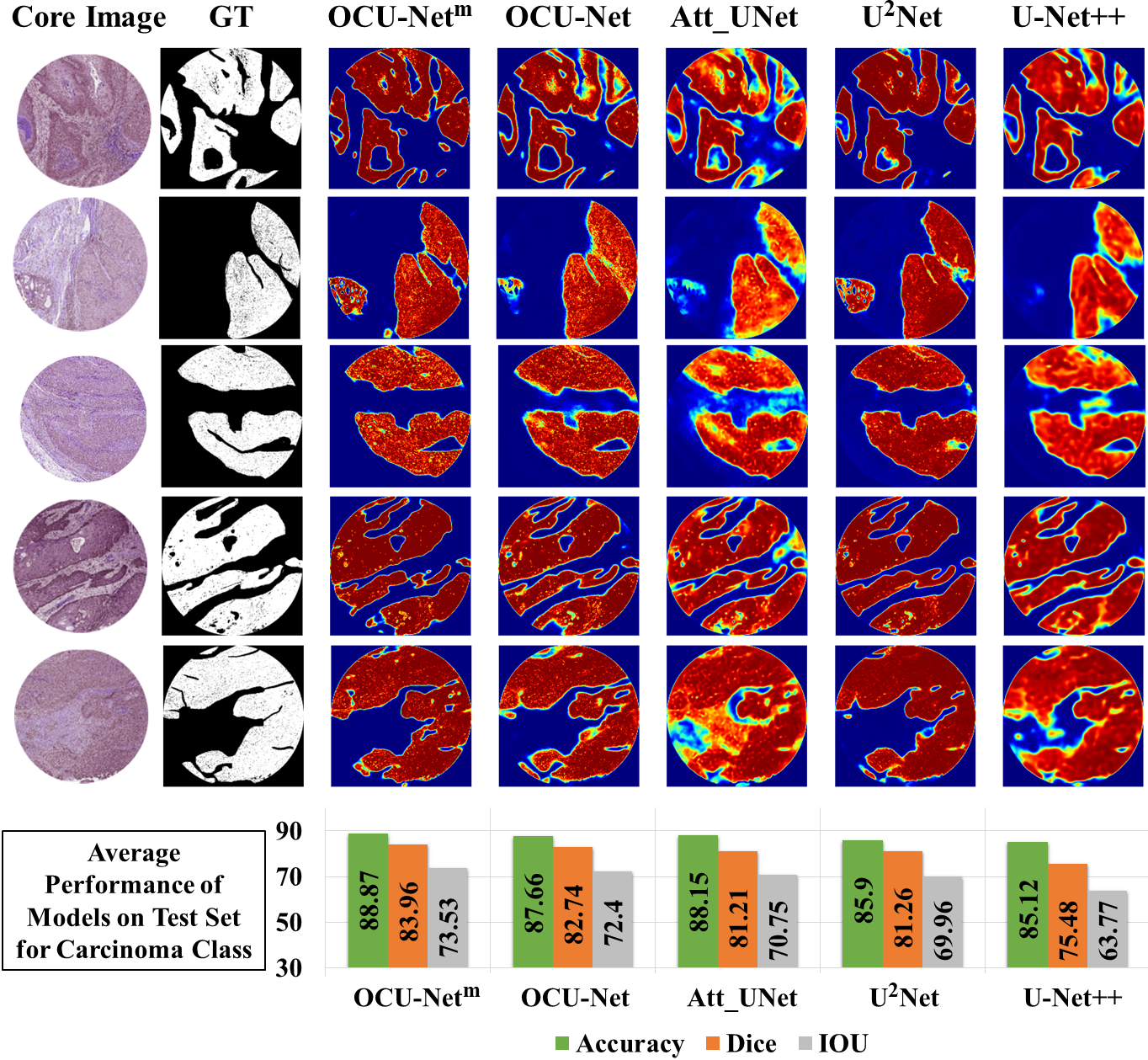}
    \caption{Evaluation of Carcinoma class segmentation performance using various models on the ORCA dataset for oral cancer detection. Classes are represented by: White - Carcinoma pixels, Black - combining both: Non-carcinoma tissue pixels and Non-tissue pixels. The bar chart presents the average performance of each model on the testing-set specifically for the Carcinoma class segmentation.}
    \label{fig:orca-results}
\end{figure*}

\subsection{Visual Validation using the Testing Dataset}
Figure~\ref{fig:orca-results} displays attention maps (heatmaps) generated by each model specifically for the carcinoma class. These heatmaps visualize the model's ability to detect the cancer class by using a color gradient, where warmer colors, such as red or orange, represent areas with high attention values, and cooler colors, like blue or green, represent lower attention values. The purpose of these heatmaps is to provide a clear understanding of the attention power of each model in detecting the carcinoma class. It is worth noting that the ORCA dataset consists of three classes: White for carcinoma pixels, Gray for non-carcinoma tissue pixels, and Black for non-tissue pixels. In this figure, the classes are represented by: White - Carcinoma pixels, Black - combining both non-carcinoma tissue pixels and non-tissue pixels.
The models are listed in order of best to worst performance, with our two OCU-Net models demonstrating the best results among the others:
OCU-Net$^m$ (ours),
OCU-Net (ours),
Our evaluation on Att\_U-Net \citep{oktay2018attention}, 
Our evaluation on U$^2$-Net \citep{qin2020u2},
Our evaluation on U-Net++ \citep{zhou2018unet++}.
The OCU-Net$^m$, OCU-Net, and Att\_U-Net (the first three models) demonstrate significantly better carcinoma segmentation accuracy compared to, U$^2$-Net, and U-Net++ (the last two models). The latter two models predict a complete white carcinoma area, lacking the fine detail found in the ground truth (GT). In contrast, the first three models are more successful in capturing the intricate details of carcinoma segmentation, resulting in a more accurate representation.

\begin{figure*}[h!]
    \centering
    \includegraphics[scale=.3, angle=0]
    {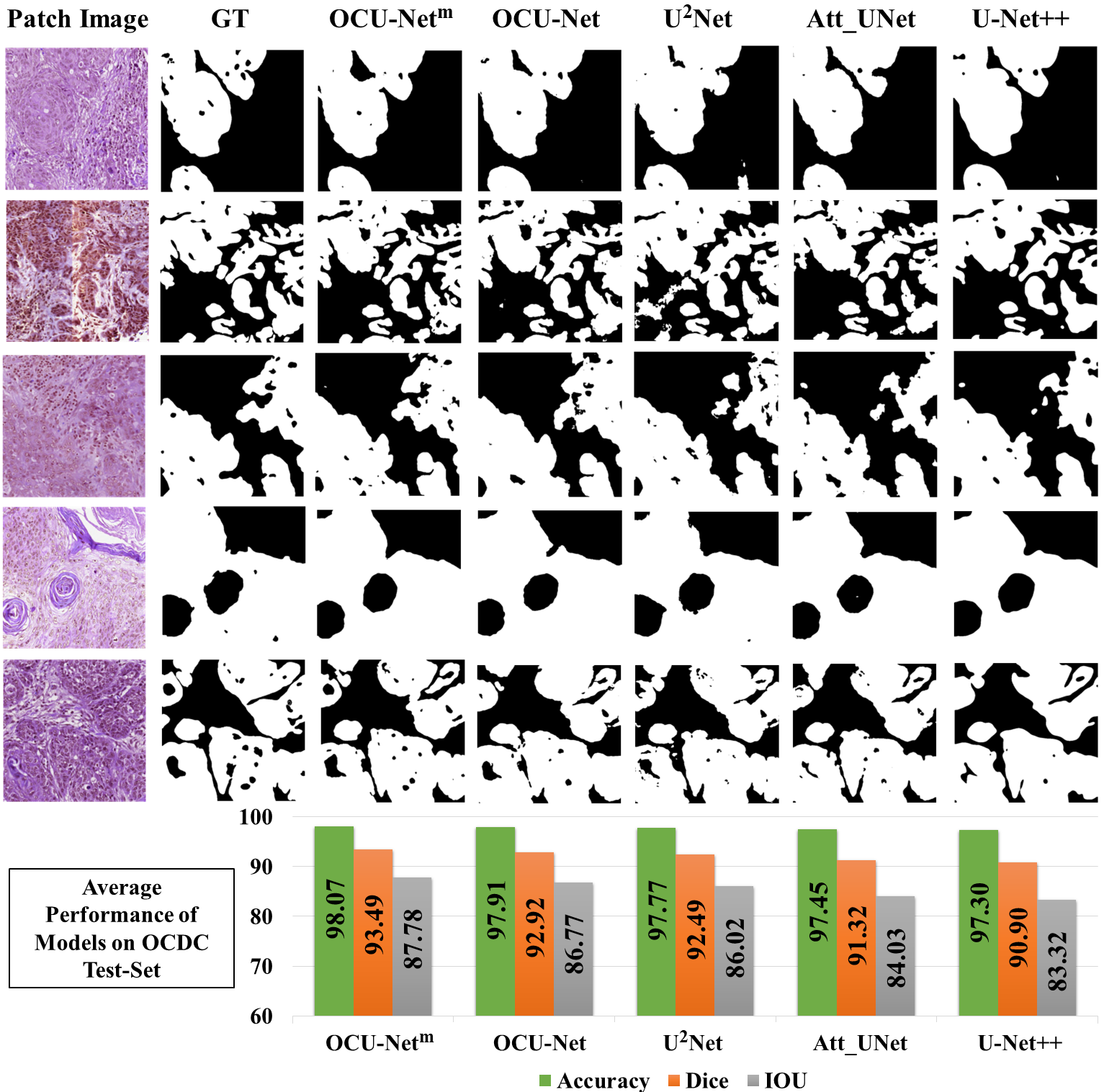}
    \caption{Segmentation performance evaluation using various models for oral cancer detection on the OCDC dataset. Classes: White - carcinoma pixels; Black - Non-carcinoma pixels. The bar chart presents the average performance of each model on the testing-set.}
    \label{fig:ocdc-results}
\end{figure*}

Figure \ref{fig:ocdc-results} displays a comparison of oral cancer patch image inputs, ground truth (GT), and detection results using various approaches on the OCDC dataset. This dataset consists of two classes: White: carcinoma pixels and Black: Non-carcinoma pixels.  The models are listed in order of best to worst performance, with our two OCU-Net models demonstrating the best results among the others:
OCU-Net$^m$ (ours),
OCU-Net (ours),
Our evaluation on U$^2$-Net \citep{qin2020u2},
Our evaluation on Att\_U-Net \citep{oktay2018attention}, 
Our evaluation on U-Net++ \citep{zhou2018unet++}.

\section{Discussion}

Oral cancer can significantly impact patients' physical, emotional, and social well-being, causing pain, difficulty eating and speaking, disfigurement, anxiety, and depression 
\citep{jehn2019physical}.
Accurate diagnosis of oral cancer is critical for effective treatment and the best possible outcomes. Although biopsy is the most definitive way of diagnosing oral cancer, its diagnosis can be subjective and affected by pathologists' experience and fatigue, leading to delayed or incorrect diagnoses. New technologies, such as deep learning, are being developed to reduce subjectivity, increase accuracy, and decrease the workload of pathologists, improving the overall diagnostic process 
\citep{das2020automated}.

Our OCU-Net architecture presents an innovative design, incorporating unique components specifically tailored for oral cancer detection in H\&E images. This innovative network integrates elements such as the novel channel and spatial attention fusion (CSAF) module, residual blocks inspired by \citep{zhou2018unet++, ibtehaz2020multiresunet, yu2021tumor}, multi-scale fusion similar to \citep{zhou2018unet++, ibtehaz2020multiresunet}, ASPP modules \citep{chen2017deeplab}, and SE blocks \citep{hu2018squeeze}, ultimately bolstering segmentation performance. By employing a tailored U-Net structure with symmetric encoder-decoder topology, efficient information flow, and precise localization are achieved. The CSAF module emphasizes crucial channels and spatial regions to explore contextual information, while residual blocks bridge the semantic gap between the encoder and decoder. Multi-scale fusion integrates features from different scales, the ASPP module captures multi-scale contextual data, and SE blocks selectively emphasize informative features in the channels. Leveraging pre-trained models like MobileNet-V2 \citep{singh2019shunt} as encoders further enhances the proposed OCU-Net architecture's performance and efficiency. In summary, our OCU-Net exhibits exceptional accuracy and efficacy, establishing it as a valuable resource for oral cancer detection among researchers and clinicians.

While our OCU-Net has shown significant advancements in oral cancer segmentation using H\&E images, limitations and potential areas for improvement warrant further exploration. First, acquiring diverse, larger annotated datasets is crucial for enhancing the model's generalization and robustness. Collaborating with medical institutions and experts can help accumulate extensive datasets covering various cancer stages and histological subtypes. GANs can synthesize artificial data, enriching the model, while semi-automatic data annotation tools can streamline the annotation process. Second, integrating additional features or complementary data modalities, such as molecular or genetic information, can provide a more comprehensive understanding of oral cancer detection and prognosis, enhancing OCU-Net's performance. Third, exploring hyperspectral imaging and in-depth detection techniques can lead to more effective cancer cell identification, especially in lower surface areas and underlying regions, by combining image classification and segmentation. Fourth, optimizing the model's architecture and computational complexity is vital for real-time segmentation, benefiting clinical settings. Novel model compression techniques, pruning strategies, and hardware acceleration can help achieve this without sacrificing accuracy. Finally, evaluating OCU-Net's performance against a larger cohort of human experts can offer insights into its clinical applicability and advantages over traditional methods, enabling further refinements tailored to clinical needs.

\section{Conclusions}


Our study highlights the superiority of OCU-Net models in several aspects. First, we developed OCU-Net for oral cancer detection in H\&E image datasets. Second, We propose a new channel and spatial attention fusion (CSAF) module designed to emphasize important channel and spatial areas in H\&E images while exploring contextual information. Third, OCU-Net models integrate unique components such as residual blocks in skip-connections, multi-scale fusion, squeeze-and-excitation (SE) attention blocks, and atrous spatial pyramid pooling (ASPP) modules. Fourth, we effectively utilized the efficient ImageNet pre-trained MobileNet-V2 model in our OCU-Net$^m$ and data augmentation, to enhance performance across different oral cancer datasets and facilitate the detection of diverse cancer presentations. Lastly, through comprehensive evaluations, we illustrated exceptional segmentation performance with reduced computational complexity, emphasizing OCU-Net's ability to accurately detect cancer cells in H\&E images from OCDC and ORCA datasets. This study positions OCU-Net as a promising tool for clinical settings, promoting earlier and more accurate oral cancer detection.




\bibliographystyle{elsarticle-num}

\bibliography{references}





\end{document}